\documentstyle[12pt,epsfig]{article}
\begin{document}

\hoffset 1.5 true cm
\voffset -0.4 true cm
\topmargin 0.0in
\evensidemargin 0.0in
\oddsidemargin 0.0in
\textwidth 6.7in
\textheight 8.7in
\parskip 10 pt

\def\identity{{\rlap{1} \hskip 1.6pt \hbox{1}}}
\def\sigeff{\sigma^2_{\rm eff}}
\def\R{{\cal R}}
\def\RA{{\cal R}_A}
\def\tl{{\tilde{\lambda}}}
\def\tb{{\tilde{b}}}

\def\contract#1{\kern 6pt \mathop{\vbox{\mathsurround=0pt
\ialign{##\crcr\noalign{\kern 3pt}
\downcontractfill\crcr\noalign{\kern 5pt \nointerlineskip}
$\hfil\displaystyle{\kern -6pt #1 \kern -6pt}\hfil$\crcr}}} \kern 6pt}
\def\downcontractfill{$\mathsurround=0pt\setbox0=\hbox{$\braceld$}
\braceld\leaders\vrule height\ht0 depth0pt\hfill\bracerd$}

\newcommand{\ba}{\begin{array}}
\newcommand{\ea}{\end{array}}
\newcommand{\be}{\begin{equation}}
\newcommand{\ee}{\end{equation}}
\newcommand{\bea}{\begin{eqnarray}}
\newcommand{\eea}{\end{eqnarray}}
\newcommand{\beas}{\begin{eqnarray*}}
\newcommand{\eeas}{\end{eqnarray*}}

\begin{titlepage}

\begin{flushright}
IASSNS-HEP-98/60\\
PUPT-1802\\
hep-th/9806214
\end{flushright}

\vskip 1.2 true cm

\begin{center}
{\Large \bf Tachyons and Black Hole Horizons in Gauge Theory}
\end{center}

\vskip 0.6 cm

\begin{center}

Daniel Kabat${}^1$ and Gilad Lifschytz${}^2$

\vspace{3mm}

${}^1${\small \sl School of Natural Sciences} \\
{\small \sl Institute for Advanced Study} \\
{\small \sl Princeton, New Jersey 08540, U.S.A.} \\
\smallskip
{\small \tt kabat@ias.edu}

\vspace{3mm}

${}^2${\small \sl Department of Physics} \\
{\small \sl Joseph Henry Laboratories} \\
{\small \sl Princeton University} \\
{\small \sl Princeton, New Jersey 08544, U.S.A.} \\
\smallskip
{\small \tt gilad@puhep1.princeton.edu}

\end{center}

\vskip 0.8 cm

\begin{abstract}
\noindent
Any probe which crosses the horizon of a black hole should be
absorbed.  In M(atrix) theory, for 0-brane probes of Schwarzschild
black holes, we argue that the relevant absorption mechanism is a
tachyon instability which sets in at the horizon.  We give qualitative
arguments, and some quantitative large-$N$ calculations, in support of
this claim.  The tachyon instability provides an attractive mechanism
for infalling matter to be captured and thermalized by a Schwarzschild
black hole.
\end{abstract}

\vfill

\begin{flushleft}
June 1998
\end{flushleft}

\end{titlepage}

\section{Introduction}

It is natural to ask what M(atrix) theory \cite{BFSS} has to say about
the physics of Schwarzschild black holes.  Several groups have
investigated this subject
\cite{BFKS,KlebanovSusskind,Halyo,HorowitzMartinec,Li,DMRR,BFKS2,TsLi,BFK,OhtaZhou,LiMartinec,Lowe}.
By modeling the black hole as a Boltzmann gas of 0-branes, several key
properties such as the entropy -- mass relationship and the Hawking
temperature have been derived, up to numerical coefficients of order
unity.

But the defining property of a black hole in classical gravity is the
existence of an event horizon.  In this paper we address the question
of how a horizon manifests itself in M(atrix) theory.  One's first
thought -- that the size of the horizon is given by the range of
eigenvalues of the matrices ${\bf X}_i$ -- cannot be correct, since
this range is expected to diverge in the large $N$ limit for all
states in M(atrix) theory.

In a scattering experiment the horizon area of a black hole can be
measured from its absorption cross section.  In the context of
M(atrix) theory this reflects the fact that a probe 0-brane should be
absorbed by the black hole once it crosses the horizon.

For Schwarzschild black holes in M(atrix) theory our proposal is that
the mechanism responsible for this absorption is a tachyon
instability: the probe 0-brane sees the horizon as a critical radius,
at which strings stretched between the probe and the black hole become
tachyonic.\footnote{A horizon is a global property of spacetime, which
cannot be detected by a local probe.  More precisely our proposal is
to associate the tachyon radius with a closed trapped surface, which
we shall loosely refer to as a horizon.}  This provides a precise
prescription for identifying the horizon in M(atrix) theory.

At a qualitative level, one can see that a state in gauge theory that
gives rise to such a tachyon region will have the features expected of
a black hole.  Suppose a probe enters the tachyon region.  Then the
tachyons will start to condense, so the probe will be captured with
very high probability.  Moreover, the tachyonic modes couple to all
the other degrees of freedom which make up the initial state, and the
energy released as the tachyon condenses is available to be
redistributed among all these degrees of freedom.  Thus within the
gauge theory we see that infalling matter gets absorbed and quickly
thermalized when it enters a region with a tachyon instability.  This
is all directly analogous to the behavior expected of matter falling
into a black hole.

Although our treatment will focus on 11-dimensional Schwarzschild
black holes, we expect the discussion of tachyon instability to apply
more generally to the horizon of non-extremal D-brane black holes, as
well as to non-extremal black holes in the context of anti de Sitter
space \cite{Maldacena}.

Extremal black holes must behave somewhat differently, however.  A
supersymmetric probe of an extremal black hole should still be
absorbed when it reaches the horizon, but the mechanism cannot be a
tachyon instability.  In section 3 we will give a brief discussion of
string pair creation as a possible mechanism in the extremal case.

An outline of this paper is as follows.  In section 2 we review some
facts about Schwarzschild black holes in M(atrix) theory and
absorption mechanisms in D-brane scattering.  In section 3 we argue
that a tachyon instability, if present, would have the features
expected of a Schwarzschild horizon in M(atrix) theory.  In section 4
we study the tachyon instability for several explicit M(atrix)
configurations.  Although we cannot directly study black hole states,
the examples we consider are sufficient to show that tachyons are
indeed present in M(atrix) theory, in regions which can plausibly be
associated with Schwarzschild horizons.  Section 5 contains some
conclusions.

\section{Preliminaries}

We review some relevant results concerning Schwarzschild black holes
in M(atrix) theory and absorption mechanisms in D-brane scattering.

\subsection{Schwarzschild black holes in M(atrix) theory}

The most general object in M(atrix) theory is a collection of 0-branes
with some strings stretching between them. The only truly stable
states are the gravitons, but a black hole can be thought of as a long
lived quasi-stable state. A Schwarzschild black hole has negative heat
capacity, unlike any state in SYM.  But in terms of light cone
variables a black hole actually has a positive heat capacity if the
number of non-compact dimensions is six or larger
\cite{KlebanovSusskind}.  This makes it possible to be described using
SYM.

Another feature of gravity that is relevant for black holes is that
one can't put an arbitrary amount of energy in a bounded region.  The
upper limit for a given region is the black hole mass with the horizon
area chosen appropriately.  In M(atrix) theory we can see a similar
effect. Fix the size of the matrices at $N \times N$, and choose some
volume with typical length $L$. We want to see if there is a limit on
the SYM energy, with the restriction that the object is contained
within the designated volume. This is easy to see at the classical
level of the SYM. For the moment let us define the size to be at most
the volume occupied by the eigenvalues of the matrices (which are not
necessarily simultaneously diagonalizable).  Now take nine diagonal
matrices with some distribution of eigenvalues $-L < \lambda <
L$. This state has zero energy. The most general configuration with
these eigenvalues can be obtained by acting with different unitary
matrices on each of the original nine matrices. As the magnitude of
each entry of a unitary matrix is smaller than 1, there is a limit on
the entries of the resulting nine matrices. Then it is clear that the
energy $\sim {\rm Tr} [{\bf X}_i, {\bf X}_j]^2$ is bounded.

Currently the best understood model for Schwarzschild black holes is
at the so-called BFKS point, that is when $N \sim S$ \cite{BFKS}. At
this point the temperature of the SYM is so low that quantum
fluctuations are of the same order as thermal fluctuations. To go
beyond this point, to $N \gg S$, requires a fully quantum mechanical
treatment of the system.  Following \cite{BFKS}, one can model black
holes with entropy $\sim N$ as follows.  The basic ingredient is a
state that has some expectation value for the matrices, $B_i = <{\bf
X}_i>$ (this is referred to as the background).  The SYM energy of
this state is identified with the light-cone energy of the black hole,
and the volume of the state is the volume of the black hole (we will
come back to what we mean by volume of the state). The background and
its fluctuations (which are modeled as a gas of N particles)
contribute to the entropy and energy of the state. In fact most papers
deal only with the thermodynamics of these fluctuations.

For the entropy to come out correctly it is crucial that the
statistics of the fluctuations be that of distinguishable particles.
This issue is somewhat subtle.  If all the background matrices commute
(so that they can be simultaneously diagonalized) then there is a
residual gauge symmetry that permutes the diagonal elements, which
makes the statistics that of identical particles. In this case the
entropy is very small.  To see this take a gas of free identical
particles in $d$ dimensions.  The assumed kinematics is that each
particle has momentum $p_\perp \sim 1/R_S$ set by the uncertainty
principle and mass $m=1/R$, where $R_S$ is the Schwarzschild radius
and $R$ is the radius of the null circle.  The volume of the gas is
$V=\left(R_{S}\right)^{d}=N$ at the BFKS point. The energy of the gas
is then $E \sim N R / R_{S}^{2}$. For a gas of $N$ distinguishable
particles in $d$ space dimensions the entropy is
\begin{equation}
S=N\log \left[V\left(\frac{4\pi m E}{d N}\right)^{d/2}\right]
+\frac{d N}{2}
\end{equation}
This entropy is indeed $\sim N$. If the particles are identical then
the volume $V$ would be replaced with $V/N$ and there would be very
little entropy.

If the background matrices do not commute (which means that there is a
coherent state of strings stretched between different 0-branes) then
we no longer have the residual gauge symmetry discussed above and thus
do not have identical particles. The factor $1/N$ will not appear and
the entropy will be order $N$.  But the parameter which breaks the
statistics symmetry is continuous, and it is not entirely clear at
what value the effective statistics actually changes. For instance, if
the energy in the fluctuations is much larger than the energy of the
background then it is not clear why the above analysis should apply.

Another issue is how to measure the size of a state. Clearly it cannot
simply be the volume occupied by the 0-branes (although this will
provide an upper bound) since, according to the holographic principle,
this is expected to grow with $N$ even for a graviton \cite{BFKS}. The
most physical measure of the size of a state is through scattering
experiments.  For a black hole, in particular, it is natural to use
the absorption cross-section as a definition of the horizon size.  So
before applying this definition to M(atrix) black holes we briefly
review absorption mechanisms in D-brane scattering.

\subsection{Absorption in D-brane scattering}

In D-brane scattering there are two mechanisms which contribute to the
imaginary part of the phase shift and lead to absorption.

The first mechanism is string creation.  When two D-branes have
non-zero relative velocity, there is an amplitude to pair create two
oppositely-oriented open strings stretched between them.  Since
velocity is T-dual to an electric field on the D-brane worldvolume,
and open strings carry electric charge, this is dual to the phenomenon of
pair production in an electric field \cite{BachasPorrati}.

In the scattering amplitude, string creation arises as follows.  The
one-loop eikonal phase shift for D0 -- D0 scattering has the form ($b$
is the impact parameter and $v$ is the relative velocity)
\cite{Bachas}
\begin{equation}
\delta(b,v) \sim \int_0^\infty \frac{ds}{s} e^{-b^2s/2 \pi \alpha'}
\frac{f(s,v)}{\sin sv}
\label{del}
\end{equation}
At $sv=n \pi$ the integrand has poles which contribute an imaginary
part to the phase shift.  The outgoing wavefunction of the scattered
0-brane is multiplied by $e^{i\delta}$.  Significant string creation
is possible only for $b < \sqrt{v \alpha'}$, since for large $b$ the
imaginary part is exponentially suppressed, ${\sl Im} \, \delta \sim
e^{-b^2/v\alpha'}$. This effect can be clearly seen in the SYM
truncation of the one-loop amplitude, where the norm of the outgoing
wavefunction is
\cite{Bachas,DKPS}
\begin{equation}
\vert\psi\vert^2=\tanh^8 \left( \frac{b^2}{4\alpha' v} \right)\,.
\end{equation}

Of course for string creation to be possible there must be enough
kinetic energy in the first place.  So we see that there are two
requirements for significant string creation to be possible,
\be
\label{eq:CreationConditions}
b < 2 \sqrt{v \alpha'} \qquad \hbox{\rm and} \qquad
b < 2 \pi \alpha' {m_s \over 2 g_s} v^2\,.
\ee
The second condition is the more stringent one when the relative
velocity is small.\footnote{These are semiclassical estimates.  The
small velocity regime, where the second condition dominates, really
calls for a fully quantum mechanical treatment of the D0 -- D0 system.
Such a treatment shouldn't change the qualitative behavior that string
creation vanishes as the relative velocity goes to zero.  We thank
C.~Bachas for raising this issue.}

The second mechanism for absorption we refer to as tachyon instability
\cite{BanksSusskind}.  While there is always a term $e^{-b^2s}$ in the
integrand for the phase shift, it could happen that for large $s$ the
function $f(s,v)$ behaves like $e^{+c^2s}$.  In eikonal scattering
this happens when the oscillator ground state energy of the stretched
string is negative.  Then for $b < c$ the proper time integral for the
phase shift diverges.  But by analytically continuing one sees that
what actually happens is the phase shift acquires an imaginary part.

Once a tachyon instability develops, the tachyon field will tend to
condense.  In a sense this condensation can be viewed as string
creation, since a coherent state of stretched strings is produced (the
strings just happen to have negative energy).  But this picture of
tachyons as stretched strings can be somewhat misleading in M(atrix)
theory.  In the context of SYM calculations the tachyon instability
arises when the (mass)$^2$ matrix for the off-diagonal elements has a
negative eigenvalue.  Of course the full potential ${\rm Tr}[{\bf
X}_i,{\bf X}_j]^2$ has a quartic piece that stabilizes the tachyon
field at some finite value.  When the tachyons condense, they lower
the energy of the original configuration.  They do this by making the
off-diagonal elements larger, which naively means that additional
strings are being added to the configuration.  But after the tachyons
condense, one can rediagonalize the matrices.  One finds that the
final configuration can be reinterpreted as having fewer stretched
strings present than the initial state.  Related phenomena have been
discussed in string theory \cite{Narain,Sen}.

In view of this ambiguity, we will use the term `string creation' only
in reference to the pair production of open strings with positive
energy arising from poles in the integrand.  A point which will be
important is that, even if no tachyon is present, string creation is
enhanced when $c^2$ is positive.  This is because the requirements for
string creation (\ref{eq:CreationConditions}) are relaxed to have
the form
\be
\label{eq:CreationConditionsII}
b < \sqrt{4 \alpha' v + c^2} \qquad \hbox{\rm and} \qquad
b < 2 \pi \alpha' \left({m_s \over 2 g_s} v^2 + c\right)
\ee
Thus string creation becomes possible for a wider range of impact
parameters when $c^2$ is positive.

\section{Horizons in gauge theory}

An essential feature of classical black holes is that they have an
event horizon.  That is, there is a region of spacetime from which one
cannot escape back to future infinity.  This means that classically
once an object enters the horizon it will be absorbed with certainty.

We want to understand how this happens in gauge theory.  In principle
given the wave function of a black hole state in the gauge theory we
could compute its absorption cross-section.  This absorption could
arise from one or both of the mechanisms discussed above.  We first
discuss Schwarzschild black holes, then briefly discuss the extremal
case.

\subsection{Schwarzschild black holes}

To mimic the black hole, we expect absorption to set in at a
relatively sharp radius, which should not depend on the nature of the
probe.  In particular the absorption cross section should be
approximately independent of the incoming probe momentum.  As we will
see, this is a non-trivial requirement.

We first consider string creation as a possible mechanism, and ask
whether string creation by gas of 0-branes (treated as distinguishable
particles) can be responsible for absorption by a Schwarzschild black
hole at the BFKS point.  The criterion is that if we send in a probe
0-brane with transverse momentum $p_\perp$ at an impact parameter
smaller than the Schwarzschild radius it should be
absorbed.\footnote{At the BFKS point a single 0-brane can be used to
probe a Schwarzschild black hole without much momentum transfer.}
This depends on the mean free path of a 0-brane in the gas of 0-branes
making up the black hole. The mean free path is given by
\begin{equation}
d =\frac{1}{n \sigma}
\end{equation}
where $n$ is the number density of the gas and $\sigma$ is the
absorption cross-section.  To estimate $\sigma$ we use the string
creation conditions (\ref{eq:CreationConditions}).  If the velocity is
large then the absorption radius $b \sim \sqrt{v}$ and hence $\sigma
\sim p_\perp^4$, while at small velocity $b \sim v^2$ so $\sigma \sim
p_\perp^{16}$.  The key point is that the probability of string
creation vanishes as the relative velocity goes to zero.
\footnote{These estimates are for eleven dimensional black holes.  In
D spacetime dimensions $\sigma$ behaves as $p_\perp^{(D-3)/2}$ and
$p_\perp^{2(D-3)}$, respectively.}

The condition for absorption is that the mean free path be smaller
than the Schwarzschild radius, $d < R_S$.  Using the BFKS density $n
\sim 1/l_p^9$ this translates into a lower bound on the transverse
momentum of the probe.  For large black holes it is the $\sigma \sim
p_\perp^{16}$ estimate which controls the bound: a probe must have
\begin{equation}
p_{\perp} > (R_S)^{-1/16}
\end{equation}
in Planck units to be absorbed.  This is rather unsatisfactory.  It
would indicate that a Schwarzschild black hole with radius $10^{38}
l_p \sim 1.6 {\rm km}$ can only absorb particles with momenta larger
than $4 \times 10^{-3} m_{p} \sim 10^{16} {\rm GeV}$.

This seems to rule out pure string creation as a mechanism for
absorption by the black hole.  But one can consider the following
hybrid mechanism.  Recall from (\ref{eq:CreationConditionsII}) that,
even if no tachyon develops, string creation is enhanced when the
oscillator ground state energy of a stretched string is negative.
Basically this makes the string lighter so it is easier to pair
produce them.  One could imagine that this enhances string creation to
the point where it is responsible for absorption by the black hole.

Such an effect is very plausibly present in non-extremal M(atrix)
black holes, as our examples in section 4 will show.  But given that
negative oscillator ground state energies are present and seem to play
a significant role, it is natural to imagine that a full-blown tachyon
instability might be present.  So let us look more closely at the
tachyon instability as an absorption mechanism.

Once a probe 0-brane enters the tachyon region, the tachyon mode will
start to condense and roll down its potential.  In the process it
acquires kinetic energy.\footnote{Quite unlike the string creation
scenario where much of the initially available energy is used up in
creating the string.}  But the tachyon field is coupled non-linearly
to the other degrees of freedom making up the black hole via the ${\rm
Tr} [{\bf X}_i,{\bf X}_j]^2$ interaction, so this energy is available
for distribution among all black hole degrees of freedom.  This is the
process of absorbing and thermalizing the new matter that fell into
the black hole.

One might worry that the probe 0-brane will leave the unstable region
before the tachyon condenses. The time scale for the tachyon to start
rolling down the potential is $t_0 \sim 1/ \sqrt{-m_{tachyon}^2}$.
Unless the probe just grazes the horizon, in order not to be captured
it must have a velocity of at least $v \sim R_{s}/t_0$.  At such large
velocities our quasistatic treatment of the background isn't reliable,
although we note that at very high velocities string creation could be
an important effect.  One also might worry that the energy which is
released by tachyon condensation could eject one of the other 0-branes
from the black hole.  But this is very unlikely at large $N$, when
there are many degrees of freedom present that can share the energy.

Of course 0-branes do escape from the black hole as Hawking particles
\cite{BFK}.  It would be interesting to understand how the tachyon
instability affects this process.  Suppose a particular 0-brane is
liberated, in that the strings connecting it to the rest of the black
hole happen to fluctuate to zero (either thermally or quantum
mechanically).  If this happens inside the tachyon region, then it is
still classically almost impossible for the 0-brane to escape the
black hole.  It is as if the 0-branes have to tunnel out of the
tachyon region in order to escape.

Note that the tachyon instability does not depend sensitively on the
relative velocity of the probe and target -- it is an effective
absorption mechanism even if the probe has zero velocity.  The tachyon
instability does depend crucially on the structure of the target,
however, and it is plausible that it could distinguish graviton states
(which do not have horizons) from Schwarzschild black holes (which
do), even though holography dictates that both gravitons and black
holes in a sense grow with $N$.

We feel that these properties make a tachyon instability an attractive
mechanism for absorption by a Schwarzschild black hole.  It seems
natural to associate the tachyon region with the size of the black
hole horizon.  To make this plausible, we must show that tachyon
instabilities can arise in macroscopically large regions in M(atrix)
theory.  We will do this in section 4 through consideration of some
explicit examples.

\subsection{Extremal black holes}

When a D-brane is used to probe an extremal black hole, absorption
should still set in at the horizon.  But the physical mechanism for
absorption must be different from the Schwarzschild case, since a
(supersymmetric) probe can't develop a tachyon instability.

An explicit example of such a system was studied in
ref.~\cite{DouglasPolchinskiStrominger}.  These authors used a
D1-brane to probe an extremal black hole with three charges, D1, D5,
and momentum.  One observation they made is that the horizon should be
identified with the origin of the Coulomb branch of the probe gauge
theory\footnote{This follows from the fact that in supergravity, in a
coordinate system with the horizon at $\rho=0$, they found that the
effective action for the probe has a term $f(\rho) \left(d\rho^2 +
\rho^2 d\Omega^2\right)$.  But a term of the same form arises in the
effective action on the Coulomb branch of the gauge theory, with
$\rho=0$ interpreted as the origin of the Coulomb branch.  Note the
contrast to the non-extremal case, where the horizon is located away
from the origin of moduli space \cite{MaldacenaProbe}.}.  They studied
geodesic motion on this moduli space, and argued that certain geodesics
which fall in towards $\rho=0$ are a signal of eventual absorption by
the black hole.

Unlike Schwarzschild black holes, the mechanism responsible for
absorption at the horizon of an extremal black hole cannot be a
tachyon instability.  Instead the mechanism is more plausibly one of
string pair creation.  Consider geodesic motion of a probe towards the
origin of moduli space.  The probe must have some non-zero velocity
$v$.  Then sufficiently close to the origin string creation will
become possible and the moduli space approximation will break
down\footnote{The one-loop estimate is that this occurs at $\rho =
\sqrt{v \alpha'}$.}.  String creation gives rise to an absorptive part
of the scattering amplitude, and can lead to capture of the probe by
the black hole (in gauge theory terms a transition from the Coulomb to
the Higgs branch).  As discussed above, string pair creation has a
slow rate of thermalization, since no energy is released in the
process.  But this is consistent with the black hole being extremal.

\section{Examples of tachyon regions}

In this section we calculate the tachyon instability region (TIR) in
some explicit M(atrix) configurations.  Ideally we would perform this
calculation for states corresponding to gravitons and Schwarzschild
black holes.  But our lack of precise knowledge of these wavefunctions
forces us to turn to some simpler examples, which we nonetheless feel
point out some general features of the TIR.  Our purpose is to make it
plausible that tachyons are present in M(atrix) theory, in macroscopic
regions that can be associated with Schwarzschild black holes.

The general calculational framework is as follows.  We choose an $N
\times N$ target configuration $B_i$, and place a 0-brane probe at a
position $b_i$.  This corresponds to a background in M(atrix) theory
\[
<{\bf X}_i> = \left[ 
\begin{array}{cc}
B_i & 0 \\
0 & b_i
\end{array}
\right] \,.
\]
The covariant gauge-fixed action of the M(atrix) quantum mechanics is
\cite{ClaudsonHalpern}
\beas
{\cal S} & =& \frac{1}{g_{YM}^{2}} \int dt \, {\rm Tr}
\left( D_t {\bf X}_i D_t {\bf X}_i +
\frac{1}{2}[{\bf X}_i,{\bf X}_j]^2 
+ i \Psi_a D_t \Psi_a - \Psi_a \gamma_{ab}^{i}[{\bf X}_i,\Psi_b]\right) \\
& & \qquad \qquad
+ {\cal L}_{\hbox{\small gauge fixing}} + {\cal L}_{\hbox{\small ghost}}\,.
\eeas
Expanding the action around the background to quadratic order gives
the following (mass)$^2$ matrices for the off-diagonal degrees of
freedom connecting the probe to the target (see for example
\cite{KabatTaylor}).
\bea
\label{eq:TheMasses}
M^2_{\rm YM} & = & \left[
\begin{array}{cc}
\sum_k \left(B_k - b_k\right)^2 & -2 \partial_t \left(B_j - b_j \right)\\
\noalign{\vskip 1mm}
2 \partial_t \left(B_i - b_i \right) & \hskip 1mm  \delta_{ij} \sum_k
\left(B_k - b_k\right)^2 + 2 [B_i,B_j]
\end{array}
\right] \\
\noalign{\vskip 0.2 cm}
M^2_{\rm fermi} & = & \sum_i (B_i-b_i)^2 \otimes \identity_{16 \times 16}
+ i \partial_t \left(B_i - b_i\right) \otimes \gamma^i + {1 \over 2} [B_i,B_j]
\otimes \gamma^{ij} 
\nonumber \\
\noalign{\vskip 0.2 cm}
M^2_{\rm ghost} & = & \sum_i (B_i-b_i)^2
\nonumber
\eea

We want to identify the probe positions which give rise to at least
one negative (mass)$^2$ eigenvalue.  Clearly $M^2_{\rm ghost}$ is
positive definite so we do not need to worry about the ghosts.  As
long as the background is static $M^2_{\rm fermi}$ is also the square
of a Hermitian matrix and hence is positive definite.  So tachyons do
not arise in the fermion sector and we can ignore them as
well.\footnote{In the context of string theory this is because the
Ramond sector always has zero oscillator ground state energy.}

This leaves the Yang-Mills degrees of freedom.  If the background is
time independent then $A_0$ does not mix with ${\bf X}_i$, and we only
need to consider the following $9N \times 9N$ (mass)$^2$ matrix for
the ${\bf X}_i$ degrees of freedom.
\be
\label{eq:MassMatrix}
\left(M^2\right)_{ij} = \delta_{ij} \sum_k \left(B_k - b_k\right)^2
+ 2 [B_i,B_j]
\ee
Even if the background does depend on time, dropping $A_0$ in this way
sets a lower bound on the size of the region with tachyons: given an
$n \times n$ Hermitian matrix with smallest eigenvalue $\lambda_{\rm
min}$, deleting one row and one column of the matrix produces a new
$(n-1) \times (n-1)$ matrix whose smallest eigenvalue is larger than
$\lambda_{\rm min}$ \cite{GR}. So for the rest of the paper we will
only analyze the mass matrix (\ref{eq:MassMatrix}).

More generally, in situations where the target must be treated quantum
mechanically, we can promote (\ref{eq:MassMatrix}) to an operator
equation.  That is, we can promote the classical background $B_i$ to a
$u(N)$-valued quantum operator ${\bf X}_i$, and write
\footnote{From now on ${\bf X}_i$ will denote the $N \times N$
operator corresponding to the target, not to be confused with the full
$(N+1)\times(N+1)$ M(atrix) field.}
\be
\label{eq:QuantumMassMatrix}
\left(M^2\right)_{ij} = \delta_{ij} \sum_k \left({\bf X}_k - b_k\right)^2
+ 2 [{\bf X}_i,{\bf X}_j]
\ee
where $[\cdot,\cdot]$ denotes the commutator in $u(N)$.  A tachyon
instability is present if the expectation value $<\psi \vert M^2 \vert
\psi>$ has a negative eigenvalue.  This can be regarded as a proposal
for defining the quantum horizon radius operator for a Schwarzschild
M(atrix) black hole.

\subsection{An $N = 2$ example}

We start by considering the case $N=2$.  This is mainly a warm-up to
illustrate some properties of the TIR.  An interesting feature which
will emerge is that the TIR does not have to be connected.

We consider backgrounds of the following form, involving two 0-branes
and some stretched strings.
\begin{eqnarray}
{\bf X}_1(t) & = & a_1(t) \sigma^{3} \\
{\bf X}_2(t) & = & a_2(t) \sigma^{1}
\label{bac1}
\end{eqnarray}
The $\sigma^i$ are Pauli matrices.  In general the operators $a_1(t)$,
$a_2(t)$ obey Heisenberg equations of motion, but for illustrative
purposes we will assume that there is some classical time-independent
background
\begin{eqnarray}
B_1 & = &  a_1 \sigma^{3} \\
B_2 & = &  a_2 \sigma^{1}
\end{eqnarray}
that approximates the properties of the state we are interested in.
The constants $a_1$, $a_2$ which appear in the background can be
thought of as suitable time averages of quantum expectation values.

The mass matrix for the bosons has the form
\[
M^2 =\left(
\begin{array}{cc}
C & 2D \\
-2D & C
\end{array}
\right)
\]
where $C$ and $D$ are $2 \times 2$ matrices.
\begin{eqnarray}
C & = &(a_1^2 +a_2^2 +b_{1}^{2} +b_{2}^{2}) \identity +2a_1b_1 \sigma^3 +
2a_2b_2 \sigma^1 \\
D & = & 2ia_1 a_2\sigma^2
\end{eqnarray}
We want to find the region in $(b_1,b_2)$ space where the probe has a
tachyon instability, {\em i.e.} where $M^2$ has a negative eigenvalue.
The boundary of this region can be found by looking for zeroes of the
determinant.  Without loss of generality we assume $a_2^2>a_1^2$, and
find that the tachyon region is
\begin{eqnarray}
0\leq b_{1}^{2} \leq \sqrt{12a_1^2 a_2^2 +4(a_2^2-a_1^2)b_{2}^2}-a_2^2-
b^{2}_{2}+a_1^2 \\
{\rm max}\left(0,a_2^2-a_1^2 -\sqrt{12}a_1a_2\right) \leq b_{2}^2 \leq
a_2^2-a_1^2 +\sqrt{12}a_1a_2
\end{eqnarray}
The maximum of $b_{1}^{2}$ is 
\begin{equation}
{\rm max}\left(\frac{3a_1^2 a_2^2}{a_2^2-a_1^2},
\sqrt{12}a_1a_2-a_2^2+a_1^2\right)
\end{equation}
If $a_1=a_2$ then there is a tachyonic instability in a disk of radius
$b^2=\sqrt{12} a_1^2$.  If $a_2^2-a_1^2-\sqrt{12}a_1a_2 > 0$ then
there are two distinct regions in which there is a tachyonic
instability.  If $a_2^2 \gg a_1^2$ then the TIR consists of two
regions, each roughly a circle of radius $\sim a_1$, centered at
$(0,a_2)$ and $(0,-a_2)$.

To look more precisely at the imaginary part, let us compute the phase
shift of a 0-brane scattered from this configuration.  To simplify
things we will take the impact parameter $b$ to be in the $x^3$
direction (and not as before in directions $x^1\,x^2$), and give the
probe 0-brane velocity $v$ in direction $x^4$. In this case there is
still string creation and a tachyonic instability so this simple
example serves to illustrate the general case.  Diagonalizing the
(mass)$^2$ matrices (\ref{eq:TheMasses}) leads to the one-loop phase
shift
\bea
\label{pm}
\delta(b,v) & = & \int \frac{ds}{s} \, e^{-(b^2+a_{1}^{2}+a_{2}^{2})s}
\frac{1}{\sin sv} \, (4+2\cosh 4a_{1}a_{2}s
+2\cos 2sv \\
& & \qquad \qquad \qquad \qquad \qquad
-8\cos sv \cosh 2a_{1}a_{2}s)\nonumber
\eea
and there is a tachyonic instability for $b^2 < 4a_{1}a_{2}-a_{1}^{2}-a_{2}^{2}$.
Notice that there is no tachyon in the
fermionic sector (the last term in equation(\ref{pm})).  From
equation (\ref{pm}) the norm of the outgoing wave function of the
scattered 0-brane can be computed to give
\begin{equation}
|\psi_{out}|=\frac{(\cosh \frac{\pi r^2}{v}-\cosh \frac{\pi c}{v})^4}{8\cosh^{6}
\frac{\pi r^2}{2v}
(\cosh \frac{\pi r^2}{v}+\cosh \frac{2\pi c }{v})}
\label{no}
\end{equation}
where we have put $r^2=b^2+a_{1}^{2}+a_{2}^{2}$ and $c=2a_{1}a_{2}$.
Let us look at equation (\ref{no}). If $a_{2}=0$ then we are reduced
to scattering one 0-brane from two other 0-branes with the usual
answer of $\tanh^{8} \left(\pi b^2/v\right)$.  In the small $v$ limit
if $r^2<4a_1 a_2$
the norm of the wave function is $\sim 0$.
Furthermore, even for small $v$ and $r^2$ outside
tachyonic region, if $r^2-4a_{1}a_{2} \sim v$ then the norm of the
wave function is $\sim 1/4$ signaling the enhancement of
string creation close to (but outside) a TIR.

\subsection{Spherical membranes}

The next set of examples we consider are spherical membranes
\cite{GoldstoneHoppe,dhn,KabatTaylor,Rey}.  A round spherical
membrane of radius $r$ can be described in M(atrix) theory by setting
\be
B_i = \left\lbrace
\begin{array}{ll}
{2 \over N} r J_i & i=1,2,3 \\
0 & {\rm otherwise}
\end{array}
\right.
\label{eq:SphericalMembrane}
\ee
The sphere is centered at the origin and extends in the directions
$x^1\,x^2\,x^3$.  The matrices $J_i$ are generators of $SU(2)$ in the
$N$-dimensional representation, normalized to satisfy
\[
[J_i,J_j] = i \epsilon_{ijk} J_k \qquad
\sum_i J_i^2 = {N^2 - 1 \over 4} \, \identity\,.
\]
We take the radius of the membrane to be much larger than the Planck
length, so that this background can be treated semiclassically.  The
M(atrix) equations of motion imply that the membrane starts to
collapse.  Eventually it will fall through its Schwarzschild radius
and form a black hole.  But black hole formation seems out of reach of
the semiclassical approximation, so we will study the membrane only
when it is at its initial (maximal) radius, with ${dr \over dt} = 0$.
This is feasible at large $N$ because the time to collapse increases
linearly with $N$ (the formula is given in \cite{KabatTaylor}).

The M(atrix) background (\ref{eq:SphericalMembrane}) is a good
approximation to a well-behaved state which is present in the full
M-theory.  It corresponds to a fundamental membrane, which has a
degenerate horizon.  In accord with this, we will verify that the
region with tachyons for these matrices vanishes in the large $N$
limit.  This does provide evidence for the hypothesis that the TIR
should be associated with the size of the horizon.

We place a probe at a position $b_i$, which we take for simplicity to
be non-zero only in the first three coordinates.  Inserting the
membrane background (\ref{eq:SphericalMembrane}) into the (mass)$^2$
matrix (\ref{eq:MassMatrix}), one finds that the non-trivial
eigenvalues come from the $3N \times 3N$ block
\[
\left(M^2\right)_{ij} = \left(r^2\left(1-1/N^2\right) 
+ \vert b \vert^2\right) \delta_{ij} \identity_{N \times N}
- {4 \over N} r b \cdot J \delta_{ij}
+ i {8 r^2 \over N^2} \epsilon_{ijk} J_k
\]
where the indices $i,j,k = 1,2,3$.

Finding the exact eigenvalues of this matrix is a difficult problem.
But, at leading order in $1/N$, one can proceed as if the different
SU(2) generators commute.  Treating the $J_i$'s as if they were
numbers, and diagonalizing, one finds that the lowest eigenvalue comes
from an $N \times N$ block given by
\[
-{4 \over N} r b \cdot J - {8 r^2 \over N^2} \sqrt{J_1^2 + J_2^2 + J_3^2}\,.
\]
In turn diagonalizing this block, one finds that the smallest eigenvalue
of $M^2$ is
\[
\lambda_{\rm min} = \left(\vert b \vert - r (1-1/N)\right)^2 - 2 r^2 / N^2
\]
where we dropped terms of order $1/N^3$.  So the smallest
eigenvalue is negative if the probe is located in the region
\[
r\left(1-1/N\right) - \sqrt{2} r / N \, < \, \vert b \vert \, < \,
r\left(1-1/N\right) + \sqrt{2} r / N\,.
\]
That is, tachyons are present in a region which is centered at
roughly the radius of the sphere, $\vert b \vert \approx r$, and
extends in either direction to form a shell of thickness $\sim r/N$.
In the large $N$ limit the region with tachyons disappears.  This is
in accord with our expectations, since the fundamental membrane of
M-theory has a degenerate horizon.

The closely related example of a 0-brane probing a toroidal membrane
was studied in \cite{AharonyBerkooz,LifschytzMathur}.

\subsection{Gaussian Random Matrices}

To motivate this final example, we observe that a precise understanding of
the relationship between M(atrix) theory tachyons and Schwarzschild
black hole horizons would require finding the black hole wavefunction
$\psi_{\rm hole}({\bf X}_1,\ldots,{\bf X}_9)$.  Then we could compute the
expectation value of the (mass)$^2$ matrix
\[
<M^2> = \int d^{\,9}{\bf X} \, \psi_{\rm hole}^*({\bf X}) M^2
\psi_{\rm hole}({\bf X})
\]
and find its lowest eigenvalue as a function of the position of the
probe.  If our proposed identification is correct, the lowest
eigenvalue should become negative when the probe is exactly at the
Schwarzschild radius.

In attempting this, we are hampered by our poor understanding of the
black hole wavefunction.  Although many of its properties have been
discussed in the literature
\cite{BFKS,KlebanovSusskind,Halyo,HorowitzMartinec,Li,DMRR,BFKS2,BFK,OhtaZhou,LiMartinec},
it is not understood precisely enough to make a calculation of $<M^2>$
possible.  Nonetheless, several general properties which the
wavefunction of a Schwarzschild black hole should satisfy are clear.
\begin{enumerate}
\item
At the BFKS point the wavefunction should have support on matrices
${\bf X}_i$ whose eigenvalues fall within a finite range.  The range of the
eigenvalues is presumably of order the Schwarzschild radius.
\item
A Schwarzschild black hole is spherically symmetric, so the wavefunction should
be invariant under $SO(9)$ rotations ${\bf X}_i \rightarrow R_{ij} {\bf X}_j$.
\item
The gauge symmetry of M(atrix) theory requires that the wavefunction be
invariant under unitary transformations ${\bf X}_i \rightarrow U^\dagger
{\bf X}_i U$.
\end{enumerate}

Our strategy is to compute the distribution of eigenvalues of
$M^2$, not with respect to the true probability distribution
given by solving M(atrix) theory
\[
P({\bf X}_1,\ldots,{\bf X}_9) = \psi^*_{\rm hole}({\bf X})
\psi_{\rm hole}({\bf X}),
\]
but rather with respect to the Gaussian probability distribution
\[
P({\bf X}_1,\ldots,{\bf X}_9) = {\rm const.} \,\,\, {\rm exp}
\left(-{1 \over 2 \sigma^2} {\rm Tr} \sum_i {\bf X}_i^2\right)\,.
\]
We regard the width of the Gaussian $\sigma$ as a freely adjustable
parameter.

The Gaussian ensemble in fact satisfies the general properties
expected of a black hole listed above.  The fact that the ensemble
is $SO(9) \times U(N)$ invariant is obvious.  Also, for large $N$ the
distribution of eigenvalues of any one of the matrices ${\bf X}_i$ is
given by a Wigner semicircle
\be
\rho(\lambda) = {1 \over 2 \pi N \sigma^2} \sqrt{4 N \sigma^2 - \lambda^2}\,.
\label{eq:Wigner}
\ee
The large-$N$ limit is taken with the 't Hooft coupling $\sigeff
\equiv N \sigma^2$ held fixed, so that the eigenvalues of the matrices
${\bf X}_i$ all lie within the finite interval
\be
\label{eq:WignerRange}
-2 \sigma_{\rm eff} \le \lambda \le 2 \sigma_{\rm eff} \,.
\ee

Since the Gaussian ensemble has some key features of a
Schwarzschild black hole, we expect that studying the tachyon region
in the Gaussian ensemble will point out some of the qualitative
effects which would be present in a more realistic M(atrix) theory
calculation.  Indeed, an optimist could even hope to use the Gaussian
ensemble as a reasonable variational approximation to the true black
hole wavefunction at the BFKS point \cite{KabatLifschytz}.

The practical advantage of Gaussian ensembles is that they are
relatively easy to work with.  Although we have not managed to find
the exact region with tachyons in the Gaussian ensemble, we are able
to put strict bounds on it as follows.  Write the (mass)$^2$ matrix as
a sum of two pieces.
\beas
M^2 & = & A + B \\
A_{ij} & \equiv & \delta_{ij} \sum_k {\bf X}_k^2 + 2 [ {\bf X}_i, {\bf X}_j] \\
B_{ij} & \equiv & \delta_{ij} \left(\vert b \vert^2 - 2 \sum_k b_k {\bf X}_k\right)
\eeas

First we study the eigenvalue distribution of the matrix $A$, which corresponds to placing
the probe at the origin (setting $b=0$).  It is convenient to work with
dimensionless eigenvalues $\tl$ defined by $\tl = \lambda/\sigeff$.
We introduce the resolvent
\[
\RA(\tl) = {\sigeff \over 9 N} \, {\rm Tr} \langle {1 \over \sigeff \tl - A} \rangle\,.
\]
The normalized distribution of eigenvalues $\rho_A(\tl)$ can be obtained
from the resolvent via
\[
\rho_A(\tl) = - {1 \over \pi} {\rm Im} \, \RA(\tl + i \epsilon)
\]
as a consequence of the identity ${1 \over x + i \epsilon} = {\rm P.V.}
{1 \over x} - i \pi \delta(x)$.
In the appendix we show that the resolvent, computed at leading order in $1/N$,
satisfies the following cubic equation.
\be
73 \tl \RA^3 + 584 \RA^2 + 9 (9-\tl)\RA + 9 = 0
\label{eq:cubic}
\ee
Of the three solutions to this equation, only one has a negative
imaginary part.  Solving for $\RA$ and extracting the imaginary part
gives the eigenvalue distribution shown in Fig.~1.  As can be seen
from the graph, the lowest eigenvalue of $A$ is
\be
\label{eq:Amin}
\lambda_A^{\rm min} \approx -8.0 \sigeff \,.
\ee
The key fact is that the lowest eigenvalue is {\em negative}.

\begin{figure}
\epsfig{file=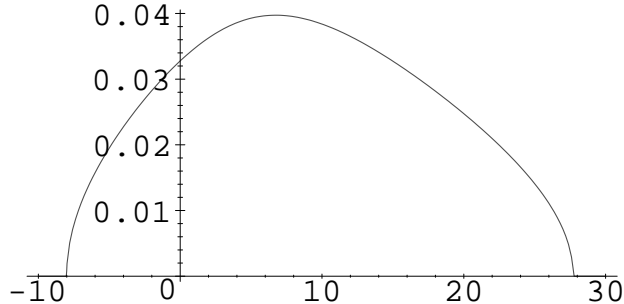}
\caption{The eigenvalue distribution $\rho_A(\tl)$.  Equivalently, the
density of stretched string states vs.~(mass)$^2$ for a probe at the center
of a Gaussian ensemble.}
\end{figure}

The distribution of eigenvalues of the matrix $B$ can be obtained starting from the Wigner semicircle
(\ref{eq:Wigner}).  It is given by
\[
\rho_B(\tl) = {1 \over 8 \pi \tb^2} \sqrt{16 \tb^2 - \left(\tl - \tb^2
\right)^2}
\]
where the dimensionless probe position $\tb$ is defined
by $\tb = \vert b \vert/\sigma_{\rm eff}$.  Note that the eigenvalues
of $B$ lie in the interval
\be
\label{eq:Brange}
\tb^2 - 4 \tb \, \leq \, \tl_B \, \leq \, \tb^2 + 4 \tb \,.
\ee

Having found the spectra of $A$ and $B$ separately, the lowest eigenvalue of
$M^2 = A + B$ can be bounded by
\be
\lambda_A^{\rm min} + \lambda_B^{\rm min} \,\leq\, \lambda_{M^2}^{\rm min}
\,\leq\, \lambda_A^{\rm min} + \lambda_B^{\rm max}\,.
\label{eq:MinEigen}
\ee
If $b=0$, so the probe is located at the origin,
then the matrix $B$ vanishes, and the eigenvalue distribution of $M^2$ is given in Fig.~1: a probe
0-brane placed at the center of a Gaussian ensemble has many tachyonic modes.  As $\vert b
\vert$ increases the eigenvalues of $M^2$ tend to increase, and
eventually a critical radius is reached at which there are no more
tachyons.  Applying the inequality (\ref{eq:MinEigen}) to the results
(\ref{eq:Amin}), (\ref{eq:Brange}) we can say that the smallest
eigenvalue of $M^2$ lies in the range
\be
\tb^2 - 4 \tb - 8.0 \,\leq\, \tl_{M^2}^{\rm min} \,\leq\,
\tb^2 + 4 \tb - 8.0\,.
\label{eq:Mmin}
\ee
It then follows that the critical
tachyon radius, at which the smallest eigenvalue of $M^2$ vanishes,
lies somewhere in the interval $1.47 < \tb_{\rm crit} < 5.47$.  Restoring units,
this reads
\[
1.47 \sigma_{\rm eff} < \vert b \vert_{\rm crit} < 5.47 \sigma_{\rm eff} \,.
\]

Recall from (\ref{eq:WignerRange}) that the largest
eigenvalue of the matrices ${\bf X}_i$ is $2 \sigma_{\rm eff}$.  So we
see that the tachyon instability sets in at roughly the same radius as
the size of the object defined by its range of eigenvalues.

\section{Conclusions}

We have argued that the horizon of a Schwarzschild black hole (or more
precisely, a trapped surface) is seen by a 0-brane probe in M(atrix)
theory as a critical radius at which a tachyon develops.  This gives
an interesting picture of the process by which such a black hole is
formed.  One starts with a collection of 0-branes located far from
each other.  With the right initial momenta they will approach each
other and some string creation will take place.  If enough strings are
created then a tachyon region will form. If the size of the tachyon
region is $\gg l_{p}$ then a macroscopic black hole has been created.
If another 0-brane approaches the black hole, it will be absorbed when
it enters the tachyon region.  In the absorption process the tachyon
rolls down its potential and, via its couplings to the other 0-branes
making up the black hole, thermalizes the infalling matter.

Since we do not have good control over the wavefunction of a
Schwarzschild black hole, we cannot conclusively show that it has a
tachyon region.  But by analyzing some examples where a tachyon
instability arises in M(atrix) theory, we showed that the TIR has
reasonable properties to be an indicator for a Schwarzschild horizon.
We gave examples which have macroscopic tachyon regions, even in the
large $N$ limit.  We also studied membrane states, where the TIR
vanishes at large $N$.

The most interesting example we analyzed was the tachyon instability
for Gaussian random matrices.  In this large-$N$ calculation we showed
that the size of the tachyon region is roughly the same as the spread
of 0-brane positions.  It is conceivable that the Gaussian ensemble is
a good variational approximation to the true wavefunction of a
Schwarzschild black hole at the BFKS point.  At least in the
non-supersymmetric case, Gaussians have been shown to give a fairly
good approximation for large-$N$ Yang-Mills quantum mechanics
\cite{EngelhardtLevit}.  The crucial test of this idea is to include
the fermions and compute the mass of a typical state in the Gaussian
ensemble \cite{KabatLifschytz}.

We expect that much of our discussion of tachyon instabilities can be
carried over to other non-extremal black holes built out of D-branes.
And given the conjectured relation between finite temperature SYM and
black holes in AdS space \cite{Maldacena}, our proposal suggests that
the phenomenon of tachyon instability can also be used as a gauge
theory indicator for the horizon of a non-extremal AdS black hole.

\vskip 1.0 cm
\centerline{\bf Acknowledgements}
We are grateful to Constantin Bachas, Misha Fogler, Anton Kapustin,
Igor Klebanov, Samir Mathur, Steve Shenker, and Lenny Susskind for
valuable discussions.  The work of DK is supported in part by the
Department of Energy under contract DE-FG02-90ER40542 and by a
Hansmann Fellowship. The work of GL is supported by the NSF under
grant PHY-91-57482.

\appendix
\section{Gaussian Random Matrices}

In this appendix we review some techniques of random matrix theory,
and derive the cubic equation (\ref{eq:cubic}) satisfied by the
resolvent $\RA$.

The standard reference on random matrix theory is the book by
Mehta \cite{Mehta}.  The methods we will use were explained to us by
Misha Fogler.  Although we are not sure of original references similar
methods appear in the literature in \cite{GreensFtns}.

To introduce the method, we start by considering a single $N \times N$
Hermitian matrix ${\bf X}$ chosen at random from the probability distribution
\[
P({\bf X}) = {\rm const.} \,\,\, \exp\left(-{1 \over 2 \sigma^2}
{\rm Tr} ({\bf X}^2)\right)\,.
\]
We are interested in finding the ensemble average of the Green's function
\[
G(\lambda) = {1 \over \lambda - {\bf X}}\,.
\]
$G(\lambda)$ can be expanded in perturbation theory, in terms of a `propagator'
$G_0 = {1 \over \lambda} \identity$ and an `interaction' ${\bf X}$, as
\[
G = G_0 + G_0 {\bf X} G_0 + G_0 {\bf X} G_0 {\bf X} G_0 + \cdots \,.
\]
To compute $<G>$ we must contract the ${\bf X}$ matrices in all
possible ways using the two-point function
\[
<{\bf X}_{ab} {\bf X}_{cd}> = \sigma^2 \delta_{ad} \delta_{bc}\,.
\]

We take the large $N$ limit holding the 't Hooft coupling $\sigeff = N
\sigma^2$ fixed.  Then the diagrams which contribute to $<G>$ at
leading order in $1/N$ are the rainbow diagrams, in which ${\bf X}$
propagators do not cross.  The same set of diagrams appear in the
solution of large-$N$ QCD in 1+1 dimensions \cite{planar}.  They can
be summed by setting
\[
<G> = G_0 + G_0 \Sigma <G>
\]
where the amputated, 1PI (with respect to $G_0$) self-energy satisfies
\beas
\Sigma & = & \contract{{\bf X} <G> {\bf X}} \\
       & = & \sigma^2 \identity_{N \times N} \, {\rm Tr} <G>
\eeas
Combining these two equations yields a quadratic equation satisfied by
the resolvent ${\cal R} \equiv {\sigma_{\rm eff} \over N} {\rm Tr}
<G>$.  Working in terms of the dimensionless quantity $\tl =
\lambda/\sigma_{\rm eff}$ the resolvent satisfies
\[
{\cal R}^2 - \tl {\cal R} + 1 = 0\,.
\]
Taking the appropriate solution to this equation, one finds that
${\cal R}(\tl)$ has a negative imaginary part for $-2 \leq \tl \leq 2$,
and that the distribution of eigenvalues of ${\bf X}$ is
\[
\rho(\tl) = - {1 \over \pi} {\rm Im} \, {\cal R}(\tl + i \epsilon)
= {1 \over 2 \pi} \sqrt{4 - \tl^2}\,.
\]
This is Wigner's semicircle distribution (\ref{eq:Wigner}), written in terms
of $\tl$.

Next we consider a slightly more involved example, which has many of
the features of the full problem.  Let ${\bf X}$ and ${\bf Y}$ be two
independent Hermitian matrices chosen at random from identical
Gaussian distributions.  To determine the eigenvalues of the
commutator $i [{\bf X},{\bf Y}]$ we introduce as before
\beas
G(\lambda) & = & {1 \over \lambda - i [{\bf X},{\bf Y}]} \\
\noalign{\vskip 2mm}
<G> & = & G_0 + G_0 \Sigma < G >
\eeas
The 1PI self-energy is given by
\beas
\Sigma & = & \contract{{\bf X} <{\bf Y} G {\bf Y}> {\bf X}}
+ \contract{{\bf Y} <{\bf X} G {\bf X} >{\bf Y}} \\
       & = & \sigma^2 \identity_{N \times N} {\rm Tr} <{\bf X} G {\bf X}>
           + \sigma^2 \identity_{N \times N} {\rm Tr} <{\bf Y} G {\bf Y}>
\eeas
The quantities appearing on the right hand side in turn satisfy
\beas
<{\bf X} G {\bf X}> & = & \contract{{\bf X} <G> {\bf X}} + \contract{{\bf X} <G>
{\bf X}} <{\bf Y} G {\bf Y}> \contract{{\bf X} <G> {\bf X}} \\
      & = & \sigma^2 \identity_{N \times N} {\rm Tr} <G>
+ \sigma^4 <{\bf Y} G {\bf Y}> \left(
{\rm Tr} <G>\right)^2 \\
\noalign{\vskip 0.2 cm}
<{\bf Y} G {\bf Y}> & = & \contract{{\bf Y} <G> {\bf Y}} + \contract{{\bf Y} <G>
{\bf Y}} <{\bf X} G {\bf X}> \contract{{\bf Y} <G> {\bf Y}} \\
      & = & \sigma^2 \identity_{N \times N} {\rm Tr} <G>
+ \sigma^4 <{\bf X} G {\bf X}> \left({\rm Tr} <G>\right)^2
\eeas
Other terms which could appear on the right hand side either vanish
identically or are subleading in $1/N$.  This system of equations can
be combined into a single cubic equation for ${\cal R} = {\sigeff
\over N} {\rm Tr} <G>$.  In terms of $\tl \equiv \lambda/\sigeff$, it
reads \footnote{Note that compared to the previous example an extra
power of $\sigma_{\rm eff}$ is present in the definitions of $\R$ and
$\tl$.}
\[
\tl \R^3 + \R^2 - \tl \R + 1 = 0\,.
\]
Choosing the appropriate solution to this equation, one finds that the
eigenvalue distribution of $i[{\bf X},{\bf Y}]$ is symmetric about zero, and is
non-vanishing only for $\tl$ between $\pm {1 \over 2} \sqrt{22 + 10
\sqrt{5}}$.

Finally, we turn to the spectrum of the matrix
\[
A_{ij} = \delta_{ij} \sum_k {\bf X}_k^2 + 2 [ {\bf X}_i, {\bf X}_j]
\]
where the ${\bf X}_i$ are independent, identically distributed $N
\times N$ Gaussian random matrices.  We let $i,j = 1,\ldots,d$ so that
$A$ is an $Nd \times Nd$ matrix.  As above we set
\beas
G(\lambda) & = & {1 \over \lambda - A} \\
\noalign{\vskip 2mm}
<G> & = & G_0 + G_0 \Sigma <G>
\eeas
where now $G_0 = {1 \over \lambda} \identity_{Nd \times Nd}$.  The self-energy is
given by
\[
\Sigma_{ij} = \contract{{\bf X}_k {\bf X}}{}_k \delta_{ij} + <\contract{A_{ik}
G_{kl} A}{}_{lj}>
\]
where the contractions in the second term mean that the two outermost
${\bf X}_i$'s which appear in $A_{ik}$, $A_{lj}$ are to be contracted.
This equation can be simplified somewhat by making use of $SO(d)$
invariance, which implies that we can set
\[
G_{ij} = {1 \over d} \delta_{ij} G_{kk} \qquad \Sigma_{ij}
= {1 \over d} \delta_{ij} \Sigma_{kk}\,.
\]
Then one has
\beas
\Sigma_{ij} & = & \contract{{\bf X}_k {\bf X}}{}_k \delta_{ij} + {1 \over d}
<\contract{A_{ik} G_{ll} A}{}_{kj}> \\
& = & N d \sigma^2 \delta_{ij} \identity_{N \times N} + {9 d - 8 \over d^2}
\sigma^2 \delta_{ij} \identity_{N \times N} {\rm Tr} <{\bf X}_k G_{ll}
{\bf X}_k>\,.
\eeas
The quantity appearing on the right hand side in turn satisfies
\beas
<{\bf X}_k G_{ll} {\bf X}_k> & = & \contract{{\bf X}_k <G_{ll}> {\bf X}}{}_k
+ \contract{{\bf X}_k <G_{lm}> A}
\kern - 5 pt \contract{\kern 5 pt {}_{mn} <G_{nl}> {\bf X}}{}_k \\
& & + \contract{{\bf X}_k <G_{lm}> <A}{}_{mn} G_{no} \contract{A_{op}> <G_{pl}>
{\bf X}}{}_k \\
& = & d \sigma^2 \identity_{N \times N} {\rm Tr} <G_{kk}> + \sigma^4
\identity_{N \times N} \left({\rm Tr} <G_{kk}>\right)^2 \\
& & + {9 d - 8 \over d^3} \sigma^4 <{\bf X}_k G_{ll} {\bf X}_k> \left({\rm Tr}
<G_{mm}>\right)^2
\eeas
There is a unique choice of contractions which
contribute at leading order in $1/N$, since the ${\bf X}$ propagators
cannot cross and the expectation value of an odd number of ${\bf
X}_i$'s vanishes.  The second equality is most easily obtained with the
help of $SO(d)$ invariance.

This set of equations can be combined into a single cubic equation
satisfied by the resolvent $\RA = {\sigeff \over N d} {\rm Tr} <G>$.
In terms of $\tilde{\lambda} = {\lambda \over \sigeff}$ the equation
reads
\[
(9 d - 8) \tl \RA^3 + (9d-8) (d-1) \RA^2 + d (d-\tl) \RA + d = 0\,.
\]
Setting $d=9$, this is the result (\ref{eq:cubic}) given in the text.


\end{document}